%% file: activity_river_avi.tex
\documentclass[sigconf]{acmart}

\AtBeginDocument{%
  \providecommand\BibTeX{{%
    \normalfont B\kern-0.5em{\scshape i\kern-0.25em b}\kern-0.8em\TeX}}}

\copyrightyear{2020}
\acmYear{2020}
\setcopyright{acmcopyright}

\acmConference[AVI '20]{International Conference on Advanced Visual Interfaces}{September 28-October 2, 2020}{Salerno, Italy}
\acmBooktitle{International Conference on Advanced Visual Interfaces (AVI '20), September 28-October 2, 2020, Salerno, Italy}
\acmPrice{15.00}
\acmDOI{10.1145/3399715.3399921}
\acmISBN{978-1-4503-7535-1/20/09}

\usepackage{xspace}
\usepackage{enumitem}

\def\plaintitle{Activity River: Visualizing Planned and Logged Personal~Activities for Reflection}

\newcommand{\dgcontext}{DG{\scriptsize context}\xspace}
\newcommand{\dgcompare}{DG{\scriptsize comparison}\xspace}
\newcommand{\dgengage}{DG{\scriptsize engagement}\xspace}
\newcommand{\dgauthor}{DG{\scriptsize authorship}\xspace}
\newcommand{\dgflex}{DG{\scriptsize flexible}\xspace}

\begin{document}


\title{\plaintitle}


\author{Bon Adriel Aseniero}
%% \authornote{Authors contributed equally to this research.}
\email{baasenie@ucalgary.ca}
% \orcid{0000-0002-2369-7803}
\affiliation{%
  \institution{University of Calgary}
%   \streetaddress{Address}
  \city{Calgary}
  \state{Alberta}
}

\author{Charles Perin}
\email{cperin@uvic.ca}
\affiliation{%
  \institution{University of Victoria}
%   \streetaddress{Address}
  \city{Victoria}
  \state{British Columbia}
}

\author{Wesley Willett}
\email{wesley.willett@ucalgary.ca}
\affiliation{%
  \institution{University of Calgary}
%   \streetaddress{Address}
  \city{Calgary}
  \state{Alberta}
}

\author{Anthony Tang}
\email{tonytang@utoronto.ca}
\affiliation{%
  \institution{University of Toronto}
%   \streetaddress{Address}
  \city{Toronto}
  \state{Ontario}
}

\author{Sheelagh Carpendale}
\email{sheelagh@sfu.ca}
\affiliation{%
  \institution{Simon Fraser University}
%   \streetaddress{Address}
  \city{Burnaby}
  \state{British Columbia}
}


%% uncomment this in the final version!!!
%% \renewcommand{\shortauthors}{Aseniero, et al.}

\input{1_abstract.tex}





%   \includegraphics[width=\textwidth]{figures/teaser_img-labeled.png}
%   \caption{Activity River is a personal visualization tool for planning, logging, and reflecting on self-defined personal activities.}
%   \Description{A screenshot of the Activity River main visualization view.}

\maketitle

%-------------------------------------------------------------------------
\input{2_introduction.tex}

%-------------------------------------------------------------------------
\input{3_background.tex}

%-------------------------------------------------------------------------
\input{4_design.tex}

%-------------------------------------------------------------------------
\input{5_activity_river_description.tex}

%-------------------------------------------------------------------------
\input{6_study.tex}

%-------------------------------------------------------------------------
\input{7_results.tex}

%-------------------------------------------------------------------------
\input{8_discussion.tex}
%-------------------------------------------------------------------------
\input{9_conclusion.tex}
%-------------------------------------------------------------------------
\input{10_acknowledgements.tex}

% BALANCE COLUMNS
\balance{}

\newpage

\bibliographystyle{ACM-Reference-Format}
\bibliography{activity_river}

% %%
% %% If your work has an appendix, this is the place to put it.
% \appendix





\end{document}

%% file: 1_abstract.tex
\begin{abstract}
We present \emph{Activity River}, a personal visualization tool which enables individuals to plan, log, and reflect on their self-defined activities. We are interested in supporting this type of reflective practice as prior work has shown that reflection can help people plan and manage their time effectively. Hence, we designed Activity River based on five design goals (visualize historical and contextual data, facilitate comparison of goals and achievements, engage viewers with delightful visuals, support authorship, and enable flexible planning and logging) which we distilled from the Information Visualization and Human--Computer Interaction literature. To explore our approach’s strengths and limitations, we conducted a qualitative study of Activity River using a role-playing method. Through this qualitative exploration, we illustrate how our participants envisioned using our visualization to perform dynamic and continuous reflection on their activities. We observed that they were able to assess their progress towards their plans and adapt to unforeseen circumstances using our tool.
\end{abstract}

%% file: 2_introduction.tex
\section{Introduction}

Self-reflection is an important process where people work to gain a clearer understanding of themselves through thoughtful introspection. Engaging in this type of reflection helps improve people's happiness and well-being~\cite{calvo_positive_2012}. Calendars, to-do lists, and diaries act as platforms for people to reflect on their planned activities, assess their progress towards goals, and to a certain extent, reassess how they can achieve these goals. Engaging in this kind of reflection is a type of deliberate, thoughtful action that forms the basis of many time management strategies~\cite{haraty_individual_2012,kamsin_personal_2012,payne_understanding_1993}. Yet, digital tools for this type of introspective reflection usually rely on simple automatically-tracked behavioural activities (like walking or sleeping) and are typically presented in business-style visualizations. Fitbit~\cite{fitbit}, for instance, tracks data via wearable sensors (usually on a watch or mobile phone). While there are benefits to this approach, such as people achieving healthy behaviour changes using activity trackers~\cite{speck_effects_2001}, the data visualizations (often simple bar and pie charts) are often insufficiently engaging for self-reflection tasks~\cite{botros_go_2016}.

\begin{figure}[t!]
  \centering
  \includegraphics[width=\linewidth]{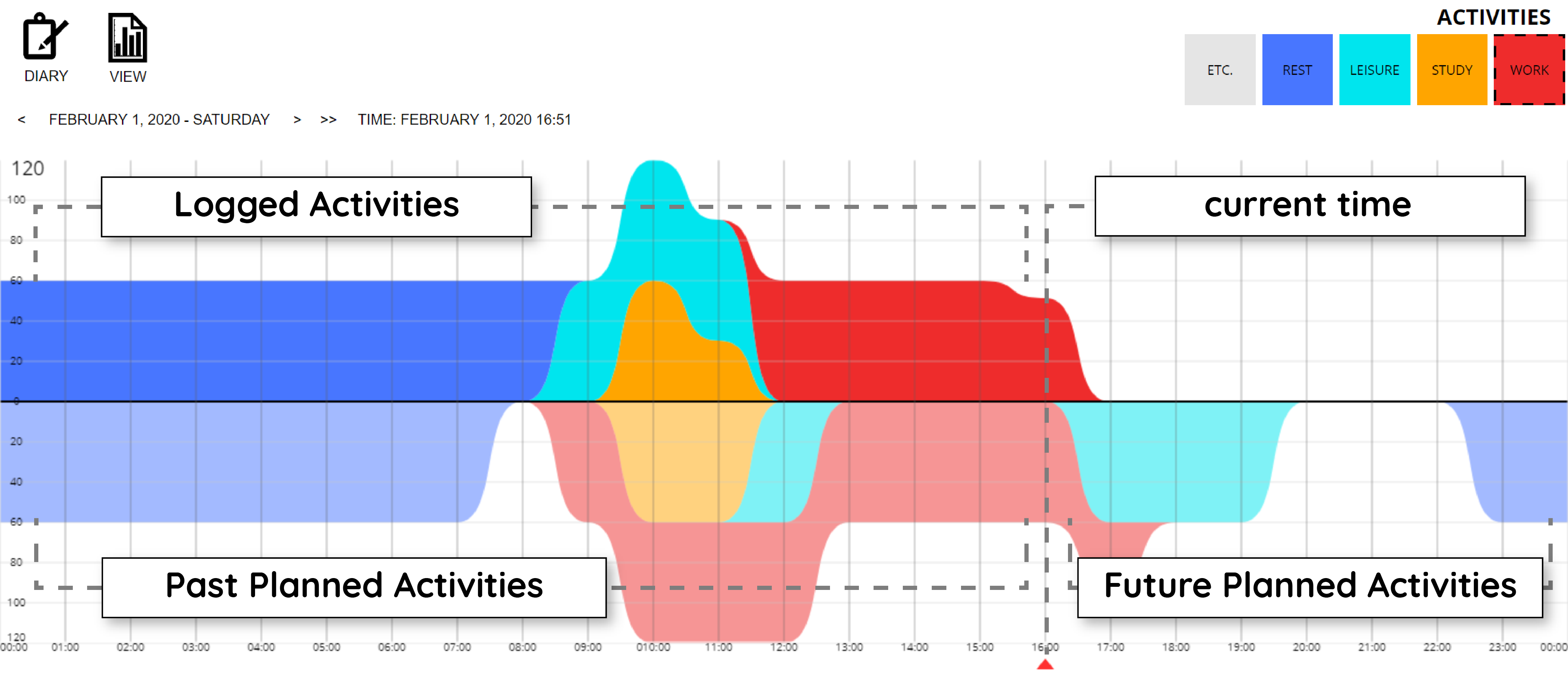}
  \caption{\label{fig:teaser}Activity River is a personal visualization tool for planning, logging, and reflecting on self-defined activities.}
  \Description{A screenshot of the Activity River main visualization view.}
\end{figure} 

As visualizations are increasingly used by people for personal reflection, it is important to learn how to design tools specifically for life-logging and introspection~\cite{huang_personal_2015}.
As of yet, we do not have a clear set of design goals for tackling this problem, and consequently, we do not have a deep understanding of how this process can be supported through visualization. Specifically, we need to understand \textit{how to support individuals as they plan and log their daily activities,} and \textit{how visual exploration can enable reflection on this type of data.}

To help shed light on these problems, we gathered ideas about planning, logging, and reflection from a combination of Information Visualization (Infovis) and Human--Computer Interaction (HCI) literature. From this literature, we identified five design goals:
\begin{enumerate}
    \item Visualize historical and contextual data (\dgcontext); 
    \item Facilitate comparison of goals and achievements (\dgcompare); 
    \item Engage viewers with delightful visuals (\dgengage); 
    \item Support authorship (\dgauthor); and 
    \item Enable flexible planning and logging (\dgflex).
\end{enumerate}
Based on these goals, we designed a visualization tool called \emph{Activity River} (\autoref{fig:teaser}) which people can use to plan, log, and reflect on their personal data. To  understand the utility of our design approach, we conducted a role-play based design study which examined how Activity River could be used to enable this kind of reflection. Results from this study indicate that Activity River supports a variety of visual planning patterns (including forward shifting, backward shifting, replacing, adding, lengthening, and shortening of activities). By enabling individuals to see these patterns, our tool allowed participants to engage in ongoing, continuous planning and reflection. This helped them commit to their planned activities and helped them adapt to changes or unexpected events. 

We make three contributions in this work: first, a set of design goals for designing visualizations for self-reflection; second,
a proof-of-concept visualization tool (Activity River) based on these design goals; and third, a role-playing exploration of this self-reflection tool and a discussion of its results.

%% file: 3_background.tex
\section{Background}

To set the stage for this work, we first touch on how the HCI community has written about self-reflection. We then discuss how personal informatics (PI) tools have been designed, as well as how personal visualizations and artistic visualizations can be used to engage people with their data, before finally turning to the role of authorship in life-logging.

Reflection occurs when individuals try to make sense of their data and lives~\cite{li_stage-based_2010}. Baumer et al. describe reflection as the process of reviewing and bringing together previous experiences and events in a manner that helps one gain insight~\cite{baumer_reflective_2015}. This concept of reflection as a key part of how we learn and gain knowledge is ubiquitous across research domains such as PI and education~\cite{korthagen_levels_2005}. While this prior work notes that people reflect for a variety of reasons, we are most interested in self-reflection---the process by which individuals learn about themselves and gain insights for self-improvement~\cite{li_understanding_2011} or behaviour change.

\subsection{Reflection and Historical Data}

Traditional ``self-tracking tools'' (like diaries, calendars, and to-do lists) contain rich contextual information about people's daily experiences which provide opportunities for deeper self-reflection. Keeping a diary, beyond the artefact itself, is a psychological process that involves putting together what we remember, what we perceive, and what we anticipate for the future. The diary author reflects on their daily activities, considers how these fit with their self-identity, and plans activities to address unmet goals. Over time, the collected writings represent an evolving account of their self-identity. Diaries can help expedite the process of constructing what McAdams describes as the ``narrative of the self'' or ``a special kind of story that each of us naturally constructs to bring together the different parts of ourselves and lives into a purposeful and convincing whole.'' \cite[p.~12]{mcadams_stories_1993}

While this implies that reflection is a long-term process, the  model of lived informatics suggests that reflection also occurs alongside data tracking, incorporating both short and long-term goals \cite{rooksby_personal_2014,epstein_lived_2015}. This can be observed in the practice of keeping calendars and to-do lists, which involves recording data that can inform individuals about their day-to-day activities \cite{haraty_individual_2012,kamsin_personal_2012}. In a study of calendar use, Payne found that calendars support \textit{prospective memory}---remembering future events \cite{payne_understanding_1993}. Payne proposed that calendars engage people's prospective memory by first helping them to set the intention to perform an action in the future, then helping them to recall the intention, and finally, in so doing, helping them to plan how to realize the action. He also found that a small majority of participants in his study also used calendars as an archive for report generation. Archived reports support \textit{retrospective memory}, or recalling past events. 

To support reflection, Li et al. suggested that designers must account for the two phases of reflection: \textit{discovery}---where individuals learn about their behavioural influences, and \textit{maintenance}---where individuals keep up with their goals. These phases rely on informing individuals of their status, history, goals, discrepancies, context, and factors affecting their behaviours \cite{li_understanding_2011}. Current PI tools provide individuals with some information such as status and history. However, most do not provide the contextual information that is necessary for people to reflect and understand their behaviours~\cite{choe_understanding_2017}. 
For example, Fitbit's dashboard provides a timeline where individuals can see their activity levels, but it does not provide context on why certain hours have more activity than others which is important if an individual wants to know how to adjust their activity.

Other tools like Toggl~\cite{toggl} are designed to support more manual activity logging. These tools enable individuals to track specific tasks and the time they spend on them. Through this contextual logging, these tools provide reports that individuals can use to examine their time use and productivity. However, such tools require diligent manual data-logging and do not integrate well with planning tools such as calendars. 

\textbf{Design challenge:} Self-reflection tools should present historical data with an appropriate level of context.

\subsection{Personal Visualizations and Goal Setting}
Researching how visualizations can help individuals in their everyday lives has received growing attention from the visualization community, prompting the research areas of Casual Infovis \cite{pousman_casual_2007} and Personal Visualization \cite{huang_personal_2015}. 
One of the goals of personal visualizations is to help individuals reflect on personal data. For example, Huang et al. suggested the use of on-calendar visualizations to contextualize physical activity data \cite{Huang_calendar_2016}. Their tool's main view is a calendar superimposed with an area graph representing a person's activity level. Enabling data exploration with this visualization helped people gain a clearer understanding about the context of their activities. Thus, they were able to reflect and give reasons as to why their activity levels would fluctuate. Choe et al. offer a preliminary explanation for how individuals could use such visual exploration for self-reflection~\cite{choe_understanding_2017}. They found that through overviews and timeline representations, their participants were able to recall past behaviours by identifying trends and peaks in the visualization.

One factor still lacking in personal visualization research is support for goal setting and planning, where individuals assign concrete goals and plan how they can achieve them. Conway and Pleydell-Pierce suggested that knowledge from life-logged data can be used to find discrepancies between expectations and actual performance, which can then be used to determine personal goals, and in turn, create plans for achieving those goals \cite{conway_construction_2000}. In this regard, Baumer's concept of \textit{breakdown}, or violating one's expectations, can also lead to reflection. In essence, tools can help individuals reflect on what behaviours they need to change by uncovering their inaccurate assumptions about themselves based on their activities in context with their plans.

\textbf{Design Challenge:} Self-reflection tools should facilitate comparison and show discrepancies between goals and achievements.

\subsection{Delightful Visualizations}
While visualizations like maps, bar charts, scatterplots, etc. can be adapted to support individual reflection, personal visualization researchers suggest considering factors such as aesthetics, playfulness, and pleasure. Huang et al. identified several works in PI and persuasive technologies which positively affected people's behaviours through playful, non-standard visualizations~\cite{huang_personal_2015}. Such qualities may contribute to long-term use of a visualization tool.

Artistic abstractions shown in aesthetically appealing ways may be particularly appropriate for self-reflection. For example, Consolvo et al. used an artistic abstraction of flowers to represent activity and fitness levels \cite{consolvo_activity_2008}. A variety of other work also suggests that aesthetically pleasing visualizations can be more engaging \cite{cawthon_effect_2007, fan_spark_2012} and more memorable \cite{bateman_chartjunk_2010}. Furthermore, people make reliable first impressions of infographic visualizations based on aesthetics \cite{harrison_infographic_2015}. As with aesthetic computing \cite{fishwick_aesthetic_2008}, the delightfulness of a visualization can elicit an emotional response from individuals, and it can also enable social engagement through shareable artefacts \cite{thudt_visual_2015}. While we scope our work to digital forms of data abstraction, other forms may also be appropriate for supporting reflection. For example, the process of creating data physicalizations (such as a piece of jewelry representing personal data) can also help individuals effectively self-reflect \cite{thudt_self-reflection_2018}. 

\textbf{Design Challenge:} Self-reflection tools should provide delightful, pleasing, and engaging visualizations.

\subsection{Authorship and Flexibility}
Early life-logging research focused primarily on providing tools for supplementing human memory~\cite{kalnikaite_now_2010}. Consequently, these early tools were designed to automatically record data, creating an ongoing record of what has transpired in one's life. A more recent approach considers active participation from individuals to complement or as an alternative to automatic logging altogether \cite{sas_self-defining_2015,thiry_authoring_2013}. These authors emphasize the value of involving individuals in authoring their life-logs, enabling them to select meaningful events and contextualize the process, thereby enriching their records. Being able to identify personally meaningful events from a larger set of mundane ones has been shown to support positive acceptance of life-logging tools \cite{sas_self-defining_2015}. In addition, Thiry et al.'s work suggested that individuals often wish to add more personalized accounts like events with their family when recording their timelines~\cite{thiry_authoring_2013}. 

\textbf{Design Challenge:} Self-reflection tools should provide ways for individuals to author their life-logs.

%Furthermore, individuals' use of calendars varies widely \cite{haraty_individual_2012}, hence, open-endedness can give agency to individuals, letting them set their own goals and achieve them as they see fit. Giving people flexibility to determine how they plan and log can help them adapt the tool to their practices \cite{ayobi_flexible_2018}.

Individuals' use of life-logging tools vary widely and have differences in how they manage their tasks and goals. Allowing individuals to author their personal goals in addition to life-logs can give them agency. In a study about personal task management, Haraty et al. proposed adding \emph{flexibility} in task management tools to accommodate these differences~\cite{haraty_individual_2012}. This can allow individuals to set their own goals and achieve them as they see fit, which can help them adapt a new tool to their  life-logging methods~\cite{ayobi_flexible_2018}.

\textbf{Design Challenge:} Self-reflection tools should provide agency to individuals through flexible means of managing their goals.

%% file: 4_design.tex
\section{Design Goals}

The design challenges we found in the literature led us to identify the following five design goals for self-reflection tools:

\begin{enumerate}[leftmargin=0pt]
\item[] \textbf{Visualize Historical and Contextual Data (\dgcontext)}. To meet the challenge of presenting historical data and context, we provide a historical timeline or archive of their data. This lets individuals assess their current status and monitor how they change over time.

\item[] \textbf{Facilitate Goal/Achievement Comparisons (\dgcompare)}. To meet the challenge of presenting discrepancies in a meaningful way, we support visual comparisons between individuals' planned and actual activities. Specifically, we explore the use of juxtaposition (placing visualizations side-by-side for comparison) \cite{gleicher_visual_2011}.

\item[] \textbf{Engage Viewers with Delightful Visuals (\dgengage)}. To meet the challenge of providing an engaging visualization, we extend \textit{streamgraphs}---a metaphor previously shown to be aesthetically appealing and appropriate for time series data \cite{byron_stacked_2008}.

\item[] \textbf{Support Authorship (\dgauthor)}. To meet the challenge of involving individuals in authoring their life-logs, we allow individuals to self-define the activities they wish to track. This permits them to personalize the tool to fit their needs and assign personally meaningful names to their activities. In doing so, they actively engage in the encoding process which is good for their memory~\cite{conway_construction_2000}. 

\item[] \textbf{Enable Flexible Planning and Logging (\dgflex)}. To meet the challenge of providing agency to individuals, we support flexible planning and logging through two different ways of recording data.
\end{enumerate}

%% file: 5_activity_river_description.tex
\begin{figure*}[t!]
 \includegraphics[width=1.0\linewidth]{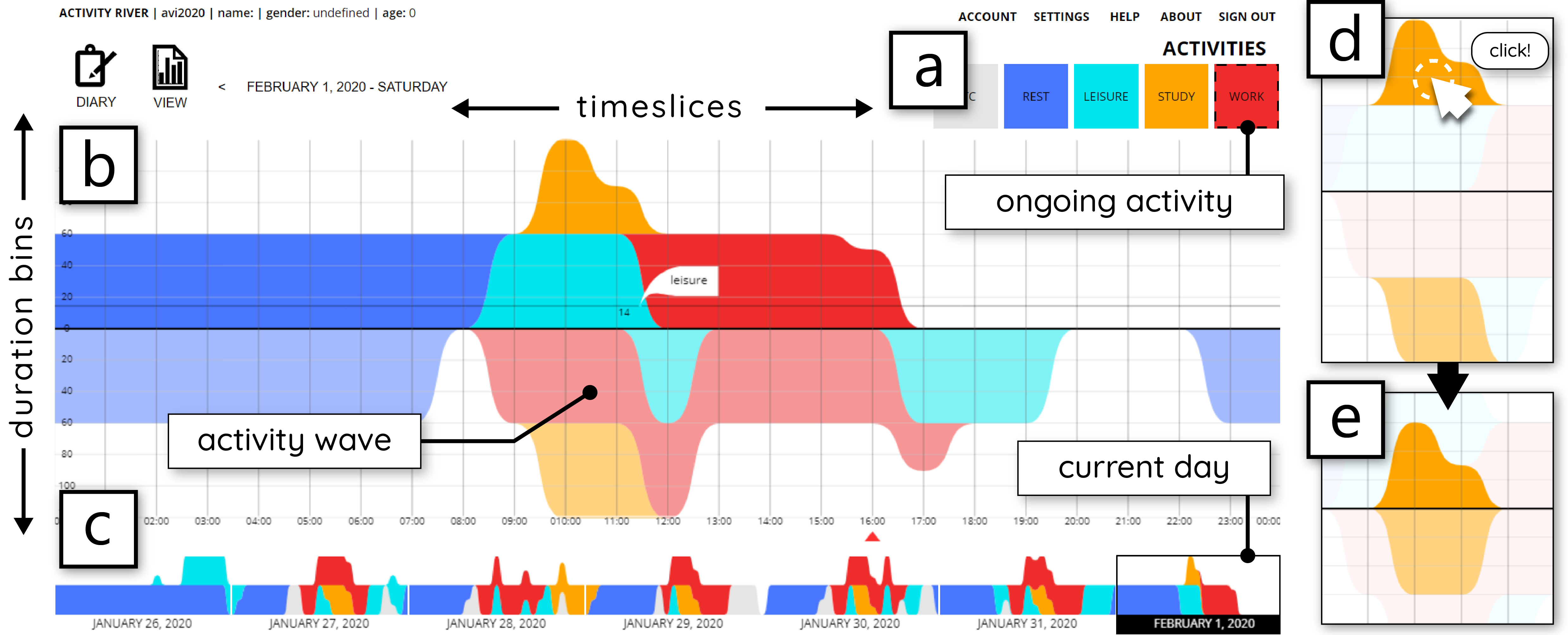}
 \centering
 \caption{\label{fig:main} The main view of Activity River: 
 (a) the legend of activities which can be toggled to log activities on-the-go, 
 (b) the timeline stream, and 
 (c) small multiples of timeline streams for the week.
 (d-e) Shows filtering and reordering interactions.
% 		\vspace{-1em}
}
\end{figure*}

\section{Activity River}
With these design goals in mind, we created \emph{Activity River} (\autoref{fig:main}), 
%an interactive tool for planning, tracking, and reflecting on time use. 
% As illustrated in Figure \ref{fig:main}, Activity River combines ideas from both calendars and diaries in a visualization that enables individuals to reflect on their past to plan their future actions. 
%Like a calendar, it allows individuals to plan future activities, and like a diary, it allows them to log which activities were actually completed. Individuals can then compare their plans against their actual activities using the timeline stream visualization. 
%We discuss each component of this proof-of-concept system in turn.
%Activity River is 
a web application that allows individuals to define, name and assign colours to the activities they wish to log. Individuals can be as precise or ambiguous as they want, depending on their goals and needs, when defining activities (\dgauthor). The tool supports open-ended methods of planning and logging activities (\dgflex). Using the diary page, individuals can plan their day by pre-selecting activities in which they wish to engage, and assigning them into time slots. Once a day is finished, they can come back to the diary and fill in the activities that actually happened. Activity River also supports an \emph{on-the-go} style of logging activities, where individuals can simply click on an activity when starting it, and then clicking it again to end it. Activities that are currently happening are signified by an animated border in the legend.

\subsection{Timeline Stream Visualization}
Our approach to address \dgcontext is to visualize data through a timeline which allows people to see their current status, and how this changes over time. We developed a visualization which we call a \emph{timeline stream} as the main visual component of Activity River. Timeline stream is a modified version of a streamgraph, where flowing stacks represent multiple time series \cite{byron_stacked_2008}. 
A timeline stream is composed of three components: activity waves, duration bins (vertical axis), and time slices (horizontal axis).
\begin{itemize}[leftmargin=0pt]
    \item[] \textbf{\textit{Activity Waves}}. Each coloured band or activity wave shows times during the day of planned or logged activities. Multiple waves can be stacked, and clicking on a wave pulls it to the baseline to ease readability (\autoref{fig:main}.d-e). 
    \item[] \textbf{\textit{Vertical Axis: Duration Bins}}. The vertical axis shows the duration (in minutes) that an individual spent on, or allotted for, a specific activity during a given hour. Bins above the horizon baseline show the amount of \textit{logged time} spent on an activity. Bins below the baseline show the \textit{planned times} for each activity. 
    \item[] \textbf{\textit{Horizontal Axis: Time Slices}}. 
    The horizontal axis contains the hours of a single day, starting at midnight on the left and continuing right. However, unlike a linear timeline \cite{aigner_visualizing_2007} where the unmarked spaces in between time markers imply progression (e.g., there are minutes implied in between 1:00 and 2:00), the time markers in the horizontal axis of the timeline stream are discrete and do not signify progression. They are akin to the discrete time markers in ThemeRiver \cite{havre_themeriver:_2000} where, for example, the space between a given year and the next does not imply continuity. 
\end{itemize}
A timeline stream combines two distinct graphs (one for planned and another for logged activities), mirrored across the central horizontal baseline. 
%The bottom stream shows planned activities, while the top stream shows actual logged activities. 
This juxtaposition allows individuals to compare their data through quick assessments of symmetry. The more symmetrical their visualization appears, the closer they are to adhering to their plan.

%% file: 6_study.tex
\section{Role-play Design Study}

To explore the strengths and limitations of our design approach, we conducted a qualitative study examining how students who use calendars and/or diaries used Activity River to plan, log, and reflect on daily activities. We focused on how they used the tool and how they behaved when adjusting and reflecting upon their plans.

\subsection{Participants}
We recruited 10 university students (six female and four male) who already life-log or use diaries, calendars, and/or other apps to plan and record their daily activities. All of them used calendars and to-do lists to a varying degree---seven had used calendars for more than a year, while the rest had used them for about a year. Only three of the participants were actively keeping diaries/journals at the time of the study, while two had kept diaries in the past. Most used such tools to organize their schedules and daily tasks. 

\subsection{Role-playing as a Study Method}
For privacy purposes, we employed a role-playing approach where we asked our participants to perform tasks using Activity River while playing the role of an imaginary student. We designed structured scenarios to guide the role-playing exercises and to encourage realistic planning, logging, and reflective behaviours. During the study, an investigator narrated and facilitated the flow of the scenario, while the participant interacted with the tool and discussed their thoughts and actions using a think-aloud method.

This approach simulated the experience of using the tool for everyday planning and reflection. 
% An alternative strategy might have been to deploy the tool for 
% participants to use; however, this 
%had the drawback of not being well-integrated into their existing tools %(e.g. calendar entries would need to be imported), and 
% might be overly disruptive to our participant's daily lives. Instead, 
Role-playing allowed participants to engage in a wide variety of behaviours---including planning and altering schedules in response to changing events---which otherwise might not happen in a given time-span. Thus, our approach can be thought of as a discount method that can supplement or lay the groundwork for more involved longitudinal studies \cite{harrison_tracking_2014,simsarian_take_2003}. Moreover, when participants role-play a character that they can relate to, they may be more open to discuss situations that they would otherwise withhold for privacy reasons.

\subsection{Procedure}
Participants completed the study individually, with each session running for about an hour. We asked them preliminary questions about their demographics, the tools they use to plan and/or log their daily activities, how they use those tools, and the types of activities they track. We then introduced them to Activity River and explained how to read the timeline stream visualization. 

\subsubsection{Characters}
We asked each participant to choose one of three characters to role-play. These characters were based on three student personas: (1) the studious senior student, (2) the active student-athlete, and (3) the carefree freshman. We made each of these characters as believable as possible, giving each a background profile detailing their goals as well as a course schedule based on our local university's calendar. We also gave each character a set of activities and asked the participants to colour code them in Activity River. We gave each participant a character information sheet that they could refer to for the duration of the study. 

\subsubsection{Scenarios}
After the participant had chosen a character, we started the roleplay with a scenario specific to their character.
Scenarios presented a typical day in life of the participant's chosen character (from when they woke up until they went to sleep). At the beginning of each day, we narrated the premise of the scenario from a script and asked the participant to use Activity River to plan their activities. After this, we asked them to set a time to wake up and begin their day. We then walked them through the day's scenario, and asked them to log their activities or browse their current status as they saw fit. We simulated the passage of time by letting the participants advance their game time in increments of five minutes or an hour.

\begin{figure}[hb]
 \includegraphics[width=\linewidth]{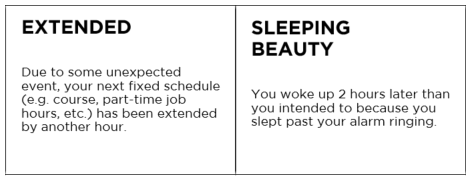}
 \centering
 \caption{\label{fig:cards}Examples of life cards. Each card varied on how long they interrupt the participant's schedule, ranging from small disruptions (such as a 1-hour delay) to big ones (like the cancellation of a whole event).}
		\vspace{-1em}
\end{figure}

Each scenario was composed of structured scenes. For every hour in the scenario, we established a scene by stating the time of day, what the character was currently doing (including any scheduled and/or on-going events), and asked the participant about their intended actions.
Once the participant had decided on an action, we asked them to simulate logging their character's activity for the hour using Activity River. We repeated this structure until the scenario for a whole day is over. To mimic the unpredictability of real life, at three specific times in the scenario (their wake-up time, around noon, and in the evening) we asked the participants to draw a card from a deck of eight \emph{life cards} (\autoref{fig:cards}). Life cards simulated unexpected real life events that would necessitate schedule adjustments. These pushed participants to adapt their schedules on-the-fly, much as they would in real life.

After the scenario, we interviewed each participant. We asked them how related they felt to their chosen character, and how knowing the character's goals and motivation affected the way they role-played and planned for the day's events. These interviews gave us insights into how participants used Activity River to reflect on their data and how they might use it in their personal lives. We also asked them how the features of Activity River compared to or contrasted with the tools they currently use.
We video-recorded each session and collected participants' comments during the interviews. We also recorded their usage of Activity River. We then used open coding to identify themes and trends in the data.

%% file: 7_results.tex
\section{Results}

\begin{figure}[t!]
 \includegraphics[width=\linewidth]{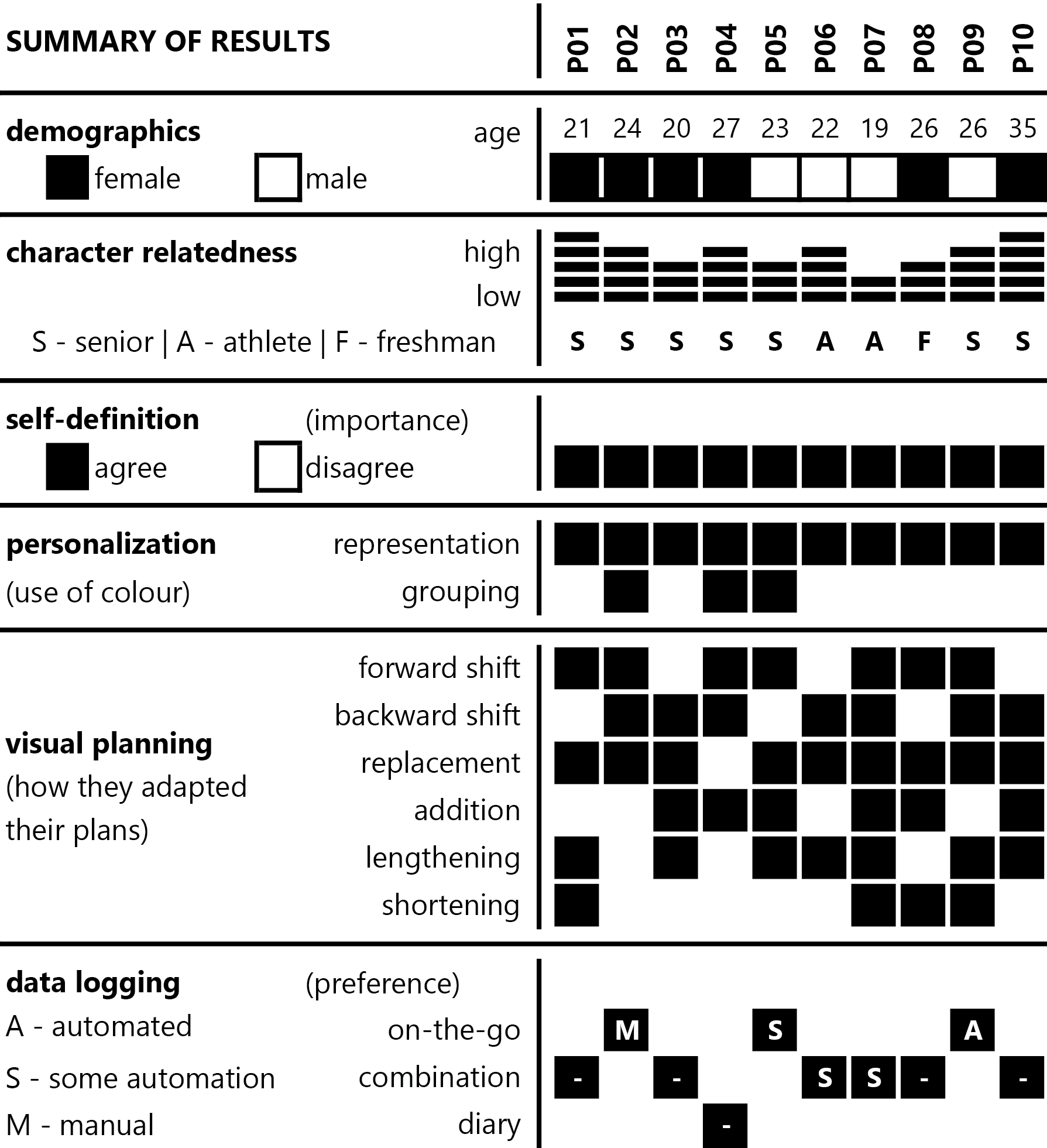}
 \centering
 \caption{\label{fig:results}Shows participants as columns. From the top, it shows their demographics, followed by how related they felt to their characters. In rows ``personalization'' and ``visual planning,'' black squares denote patterns we observed in a participant. For ``data logging,'' black squares denote their preferred logging method and levels of automation.
% 		\vspace{-1em}
}
\end{figure}

Our participants empathized readily with the characters they chose. While they varied in how closely related they felt towards their character (see \autoref{fig:results}: character relatedness), they all reported that they clearly understood the character's motivations. Participants also reported doing their best to stay true to their character while completing the scenarios. For example, while Participant 6 felt that his chosen character (the student-athlete) had some motivations which were in conflict with his own, he still stated that \textit{``[the character profile] affected me in the sense that I tried to think like the character. I know what I need to do to excel in sports so even if I had a headache, I still went to the training sessions and skipped a course, which I would normally not do myself.''} Figure \ref{fig:results} provides a summary of our results, including participants' demographics, the extent to which they felt related to their characters (scale of 0-5), and  whether they thought of self-defining activities as important or not. We also report how participants used colour to personalize their visualization (whether they assigned colours to activities based on their perceived meaning of the colour, or if they used colours to group activities together), the types of visual planning we observed in each of them, and their data-logging preferences.

%In detail, we first present six visual planning patterns performed by our participants, followed by our observations on how they gained insights while using Activity River. 

\subsection{Visual Planning}
During the planning task at the beginning of the day, all participants planned for the activities that were on their character's fixed schedule (including courses that they needed to attend). All participants also looked at their free times and started filling them with other activities that they thought were useful for achieving their character's goals. They also used the timeline stream to identify plans they could change to get closer to their character's goals.

As they went through the scenarios, all participants tried to adhere to their plans. They all used the visualization showing their planned activities as a reminder of what they still needed to do and when they planned on doing it. All participants agreed that seeing their plans visually helped them stay on track as their scenarios unfolded. Half of the participants claimed that they would have most likely overestimated the time it took to finish some activities if they were not able to see their plans.

\subsubsection{Continuous Reflection}

We observed that the timeline stream visualization enabled our participants to commit to their plans. By visualizing plans alongside what they had already done, Activity River improved participants' awareness of their schedules (\dgcontext and \dgcompare). For instance, Participant 2 reported that during her scenario, she actively used the lower portion of the timeline (the planned activities) to ensure she reached her character’s goals. When participants received life cards which interfered with their plans, they used the tool to visually identify activities which they could shift or compromise, while still feeling that they were achieving most of what they planned. This shows that they were able to perform quick bursts of decision-making to adapt to changes. 
Moreover, our interviews revealed that participants felt the visualization helped them understand what they needed to change both when dealing with immediate interruptions and when planning their long-term goals.
Broadly, participants demonstrated continuous reflection, enabled by their visual data exploration, which allowed them to make more informed modifications to their plans. In the following subsection, we highlight six visual exploration patterns that participants used as part of their continuous reflection.

\subsubsection{Modifying Plans using the Timeline Visualization}
\begin{figure}[t!]
	\includegraphics[width=\linewidth]{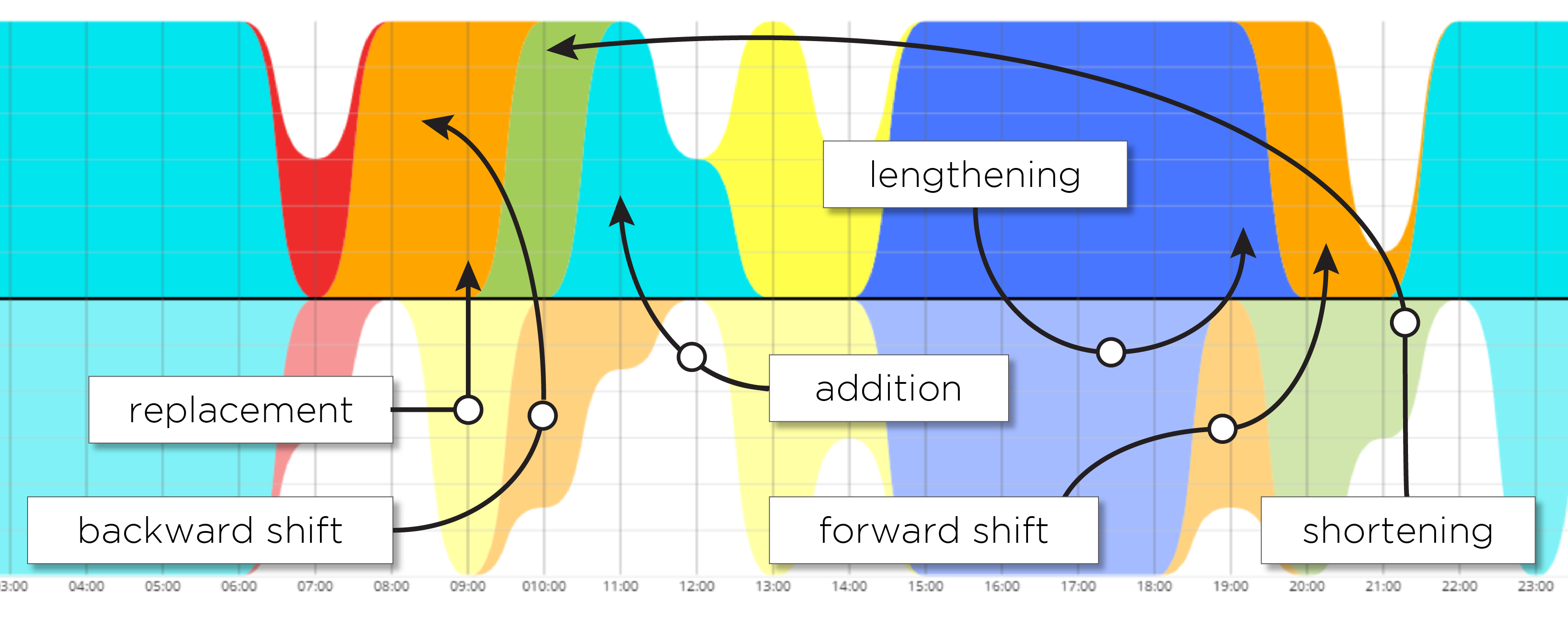}
	\centering
	\vspace{-1em}
	\caption{\label{fig:patterns}Shows the timeline stream of Participant 7 whom we observed to have used all six patterns. The labels correspond to each pattern, pointing first to where an activity was planned (circle dot) and then to where the activity actually happened, or to what replaced it (arrow head).
% 		\vspace{-1em}
	}
\end{figure}

We observed six visual planning patterns our participants used to modify their schedules when their plans were interrupted (\autoref{fig:patterns}). Participants used the timeline stream to visually search for flexible activities which they could shift to accommodate the change.

\begin{itemize}[label={},leftmargin=0pt]\label{list:visualpatterns}
	\item \textbf{Pattern 1. Forward Shift} -- We observed seven participants shifting activities to a later point in time. For example, when Participant 7 had to extend his sports activity, he pushed his studying to an hour later. This could also occur when a participant missed an activity they wanted to do, in which case they did it at a later time.
	
	\item \textbf{Pattern 2. Backward Shift} -- Seven participants shifted activities to an earlier time than they originally planned. This occurred when participants realized they had free time earlier in the day. For example, Participant 3 had one of their plans cancelled (due to a life card) so they instead performed an activity which they originally planned to do much later that day.
	
	%For example, when a morning course was cancelled, Participant 7 decided to study instead of doing it at a later time which he first planned.
	
	\item \textbf{Pattern 3. Replacement} -- Nine participants replaced at least one planned activity with a different activity. 
	
	\item \textbf{Pattern 4. Addition} -- Six participants performed an activity which they did not originally plan to do, usually during their free time. For example, Participant 4 decided to take a nap during a break in their schedule.
	
	%For example, Participant 7 caught a cold near midday, so he decided to sleep, filling an otherwise empty time slot.
	
	\item \textbf{Pattern 5. Lengthening} -- Seven participants extended the duration of an activity beyond what they had originally planned. These extensions were often compensations for missing part of an earlier instance of the activity.
	
	\item \textbf{Pattern 6. Shortening} -- Four participants cut a scheduled ongoing activity short. For example, Participant 8 planned to study for 1.5 hours but decided to shorten it to 1 hour to get more sleep.
	
	%For example, Participant 7 decided to take his leisure time earlier during the day. However, because he caught a cold, instead of having leisure for an hour and a half as planned, he only ended up having it for an hour.
	
\end{itemize}

\subsection{Drawing Insights from the Data}
Beyond using Activity River to simply make short-term decisions, participants used the visualization to uncover insights about their characters based on their timeline streams. All participants used the timeline streams to make decisions about what they should change in order to get closer to their character's goals. We observed two specific instances when the visualization triggered seemingly intuitive insights in our participants. By seeing the comparison between their planned and actual activities, they made unprompted personal connections between their character's data and themselves. For instance, after seeing her activities at the end of the scenario, Participant 10 said \textit{``[As this character,] I think I need to do more studying and working out because I [was not able to do any]. I need to invest on `me' and prioritize myself. That's the best investment for life.''} Similarly, Participant 6 said \textit{``I think my plan [as this character] shouldn't be too strict. I need to add flexible times in-between activities''} after getting a disruptive life card that required him to alter his schedule extensively. This is reminiscent of \textit{non-rational reflection} in which environmental cues trigger creative interpretations \cite{korthagen_two_1993}. 

In comparison to their methods of activity planning and logging, six participants appreciated how Activity River visualized their entire day at-a-glance so they know exactly what had happened. Those who use calendars and to-do lists said that they would like to be able to use Activity River along with their tools because it would help them know whether they were able to complete the activities that they had planned to do. For example, Participant 6 mentioned that Activity River was more motivating than his calendar because it gave him insights on whether he was achieving his planned activities or not. Participant 4 said \textit{``I use Google Calendar, but that doesn't really visualize your day. It also doesn't tell you whether you actually did or did not perform an activity. I actually come back to my Google calendar right now and edit/delete activities that I didn't get to.''} On a similar note, Participant 3 appreciated the fact that she could see the amount of time she needed for an activity. She also suggested this might help her allocate the right amount of time and not over-plan, as she had done before with her agenda. 

\subsection{Self-Defined Activities and Personalization}
Letting individuals define the activities they wish to track was one goal for Activity River's initial design (\dgauthor). 
% While we did not ask our participants to define their character's activities during the study, we talked to them about defining their own activities. 
While our participants defined the activities for their characters and not necessarily for themselves, we still asked them about their thoughts on self-defining activities.
All participants stated that the ability to define their own activities is important. One common rationale participants gave was a need for more or less specific categories that better matched their activities. They also acknowledged that not everyone is the same and some individuals may wish to track completely different activities from what they have been given. For example, Participant 7 felt that activities such as ``leisure'' and ``study'' were too broad. Because he wanted to know exactly what activity he was spending his time on, he envisioned using more specific names. Furthermore, Participant 4 felt that arbitrary activities such as ``study'' did not imply the same meaning as ``studying for an exam'' or ``studying for homework,'' hinting that individuals do in fact add personal meaning to the names of their activities. 

All the participants gave at least one or two activities a colour which they claimed best represented the activity. For example, Participant 5 said \textit{``I will choose this dark shade of blue to denote Sleep because it's like night.''} and Participant 6 said \textit{``Leisure will be this `lemony' colour because that is the colour of [beer].''} 
Some participants also used colour coding to group similar activities together. For example, Participant 1 said \textit{``School is blue because it makes me blue. Studying is [light blue] because it's similar to School.''} Later in the study, participants also used the overall colour tone of their streams to get a feel of whether they have accomplished their character's goals. There were also isolated cases where our participants had prior associations of importance to colour and used that idea to rank the activities according to their perceived importance. For example, Participant 9 said \textit{``Study is red, because red is a very important colour. Work is orange, not as important as red, but still important.''}

%% file: 8_discussion.tex
\section{Discussion}

Insights from our study supported and extended our initial design 
intuition. These extensions suggest new approaches to supporting people in planning, logging, and reflecting on their daily activities. In this section, we discuss interesting behaviours exhibited by our study participants and how they relate to our design goals. 
%We highlight the relevance of each goal and provide suggestions for improvement. Throughout, we emphasize the importance of considering our design goals in the design of future similar applications. 

\begin{figure}[b!]
	\includegraphics[width=\linewidth]{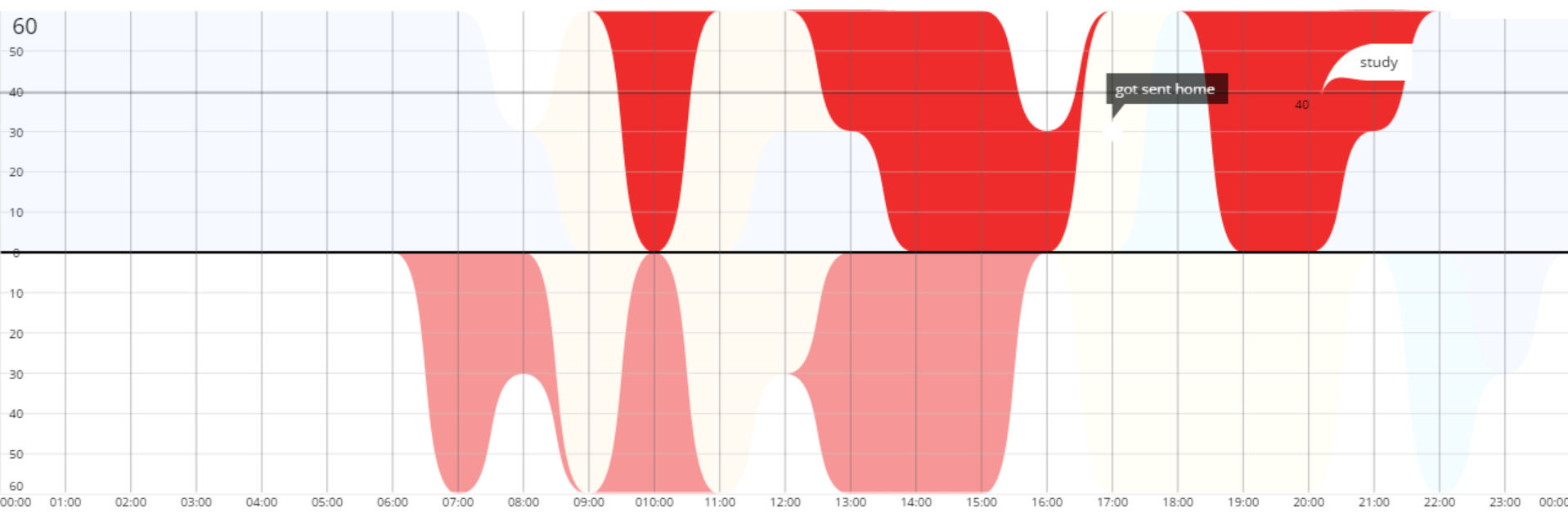}
	\centering
	\caption{\label{fig:banking}An example of a bankable activity in which the amount of time spent on an activity mattered more than when it was scheduled.
	}
\end{figure}

\subsection{Supporting Other Forms of Planning}
Activity River supports planning of activities at specific times. 
This prevents individuals from planning too many activities in a limited 
time span. However, it also pushes them to specify times for some activities for which looser specifications, such as those in a to-do list, might be more appropriate. These are activities that can be interrupted and/or have non-specific start and end times (and thus the time when participants do certain activities does not matter as long as they get to do them). These non-fixed activities are typically the ones that our participants looked for and shifted during our study when their schedules were interrupted. Participants were more concerned with the amount of time spent on such activities rather than when they happened. For example, Participant 4 planned for a total of 5.5 hours of studying, and while she did not study at the exact times she had planned, she still managed to study for the full 5.5 hours and a bit more (\autoref{fig:banking}). 

These can be thought of as \textit{bankable activities,} in which 
individuals first set an amount of time as their goal and then work 
towards that amount whenever they have time to do so. We consider bankable activities as a class of activity where identifying start times may be difficult or altogether inappropriate. For these 
activities, it is more important to show how much time has already 
been spent on the activity rather than when it happened. 
To improve our current design goal (\dgflex) and properly handle bankable activities, applications should allow individuals to plan some activities in list-form. However, to prevent over-planning, the application should also account for the duration these listed activities should take. The application could then suggest times for completing these bankable activities (for example, adding visual cues to the visualization which highlight free times). Tools could also keep ongoing counts of bankable activities and signal when the individual has met their planned duration. Nevertheless, the tool should give the individuals flexibility to plan/log bankable activities as there are times when setting a fixed time for this type of activity becomes necessary (such as when they plan to do it with another person). 

\subsection{Curiosity Sparked by Visual Aesthetics}
Our participants rated the aesthetics of Activity River highly and were interested in personally using it for more than a week in order to see how their data would look. This suggests that the aesthetics of timeline streams, with its flowing symmetry and vibrant colours, sparked curiosity in our participants. While not necessarily an emotional investment, curiosity is a motivator which can initiate further exploration of the data \cite{arnone_curiosity_2011}, and encourage continued use of a tool. While some participants needed detailed instructions to be able to read the visualization, sparking individuals' curiosity to engage with their data likely encouraged more investment in that learning process (\dgengage). Future work that evaluates the effects of visual aesthetics in tools for reflection is needed to better understand this trade-off.

\subsection{Integration with Existing Tools}
Activity River's current manual logging of data is a burden that could 
deter individuals from using it long-term. However, many people already use apps such as Toggl \cite{toggl} and Optimized \cite{optimized}, as well as diaries, to-do lists, and calendars---all of which involve manual logging of daily events at similar granularities. Although tedious, manual logging has the beneficial side-effect of providing an opportunity for people to be more involved with their data, and thus reflect on it more deeply.

The majority of our participants requested to integrate Activity 
River with tools that they already use. For instance, Participant 4 said 
she would like to connect it to her digital calendar because she wanted to see the timeline stream of her actual activities, but still wanted to 
plan using the calendar's interface. Alternatively, linking with geo-tracking tools could also allow Activity River to detect an individual's location and prompt them to record their activity. This semi-automated enhancement could help reduce the burden of manual logging \cite{li_stage-based_2010}, but still keep the person in the loop to handle the very personal task of identifying and categorizing activities. 
\subsection{Limitations}
While evaluating tools like Activity River in-the-wild is an important future step, in-the-wild studies are time-consuming and expensive. Before undergoing such a venture, approaches like our role-playing study can be used to reveal important issues and opportunities, and better ensure tools are ready for deployment. Our role-playing approach highlights how a controlled, simulated setting can still provide a sense of realism and trigger frank discussions about familiar activities. This controlled approach also reduces the risk of violating participants' privacy---which can complicate the design of real-world evaluations of personal informatics tools and may lower participants' willingness to take part.

Our choice of familiar, yet simulated activities provided us with considerable insights, and participants indicated that they saw value in the tool for themselves. However, we acknowledge that we cannot assume whether individuals would use this tool over a longer period of time. 
% We were also optimistic about how/when participants could log, and that activities of interest could be readily categorized, which may not be the case in real-life. 
Furthermore, in letting our participants choose the speed at which their scenarios advanced, they were able to always log their activities which may not be the case in real-life. 
If privacy concerns can be addressed, then we could explore how people use the tool for themselves with a field study. This could give us more insight into the utility of self-defined activities and how the tool works in real-life situations.

\subsection{Recommendations for Future Studies: Beyond Personal Activity Tracking}
In this work, we reported on how timeline streams supported six visual planning patterns and how our participants used them to manage their goals. However, these visual explorations may also be relevant for many other types of planned and logged data. Future work could investigate how timeline streams' comparative visuals could help people identify discrepancies between budgets and actual spending. For example, in examining discrepancies in software release planning budgets or in tracking medication adherence. Furthermore, we saw how our participants used Activity River to communicate and describe their thoughts (for instance, by pointing at interesting parts of the visualization). Thus, future studies could also investigate how timeline visualizations could help collaborators (managers and employees, coaches and athletes, etc.) communicate about shared planning tasks.

%% file: 9_conclusion.tex
\section{Conclusion}
Our work explored how to support self-reflection through life-logging and visualization. We designed and implemented \emph{Activity River}---a proof-of-concept tool for planning, logging, and reflecting on personal activities. 
%---based off these design goals: Visualize historical and contextual data (\dgcontext); Facilitate comparisons of goals and achievements (\dgcompare); Engage viewers with delightful visuals (\dgengage); Support authorship (\dgauthor); and Enable flexible planning and logging (\dgflex). 
In our role-playing study of Activity River, we found that our participants were able to use it to progress towards their goals. In particular, our timeline stream visualization enabled dynamic and continuous reflection, which helped individuals make quick, informed decisions, and adapt to unforeseen circumstances. We also noted that self-definition of activities and goals enriched people's life-logging experience, allowing them to tailor the application to fit their specific needs.

It is our hope that the lessons from this work will help improve future tools for activity logging. In particular, we encourage designers to consider the dual but interrelated nature of planning and reflection in future tools. Ultimately, reflecting on our design goals and findings may contribute to the design of more effective  visualization tools for self-reflection.

%% file: 10_acknowledgements.tex
\begin{acks}
This research was supported in part by: the Natural Sciences and Engineering Research Council of Canada (NSERC); SMART Technologies; and Alberta Innovates - Technology Futures (AITF).
\end{acks}